\documentclass[conference]{IEEEtran}
\IEEEoverridecommandlockouts
\usepackage{cite}
\usepackage{amsmath,amssymb,amsfonts}
\usepackage{algorithmic}
\usepackage{graphicx}

\usepackage{textcomp}
\usepackage{xcolor}

\usepackage{multirow}
\usepackage{subcaption}

\usepackage{array}

\def\BibTeX{{\rm B\kern-.05em{\sc i\kern-.025em b}\kern-.08em
    T\kern-.1667em\lower.7ex\hbox{E}\kern-.125emX}}
\begin{document}

\title{AsymLLIC: Asymmetric Lightweight Learned Image Compression\\
}

\author{
    \IEEEauthorblockN{
        Shen Wang\IEEEauthorrefmark{1},
        Zhengxue Cheng\IEEEauthorrefmark{1}\thanks{Shen Wang, Zhengxue Cheng, Donghui Feng, Guo Lu, Li Song and Wenjun Zhang are with the Department of Electrical Engineering, Shanghai Jiao Tong University, Shanghai 200240, China (e-mail: wangshen22206@sjtu.edu.cn; zxcheng@sjtu.edu.cn; faymek@sjtu.edu.cn; luguo2014@sjtu.edu.cn; song\_li@sjtu.edu.cn; zhangwenjun@sjtu.edu.cn).
        Corresponding author: Zhengxue Cheng.},
        Donghui Feng\IEEEauthorrefmark{1},
        Guo Lu\IEEEauthorrefmark{1},
        Li Song\IEEEauthorrefmark{1},
        Wenjun Zhang\IEEEauthorrefmark{1}
    }
    \IEEEauthorblockA{\IEEEauthorrefmark{1}\textit{Department of Electronic Engineering}, \textit{Shanghai Jiao Tong University}, Shanghai, China\\
    }
}

\maketitle

\begin{abstract}

Learned image compression (LIC) methods often employ symmetrical encoder and decoder architectures, evitably increasing decoding time. However, practical scenarios demand an asymmetric design, where the decoder requires low complexity to cater to diverse low-end devices, while the encoder can accommodate higher complexity to improve coding performance. In this paper, we propose an asymmetric lightweight learned image compression (AsymLLIC) architecture with a novel training scheme, enabling the gradual substitution of complex decoding modules with simpler ones. Building upon this approach, we conduct a comprehensive comparison of different decoder network structures to strike a better trade-off between complexity and compression performance. Experiment results validate the efficiency of our proposed method, which not only achieves comparable performance to VVC but also offers a lightweight decoder with only 51.47 GMACs computation and 19.65M parameters. Furthermore, this design methodology can be easily applied to any LIC models, enabling the practical deployment of LIC techniques.


\end{abstract}

\begin{IEEEkeywords}
Learned image compression, neural network, lightweight, asymmetric
\end{IEEEkeywords}
\vspace{-5pt}

\section{Introduction}
In real-world image compression scenarios, client-side computational capabilities are often limited. To achieve high compression performance while maintaining low decoding times, it becomes necessary to offload some of the computational burden to the encoding side. This approach is known as asymmetric computational architecture. 

Traditional image compression algorithms, such as JPEG\cite{wallace1991jpeg}, JPEG2000\cite{rabbani2002overview}, HEVC intra\cite{sullivan2012overview} (BPG), and VVC intra\cite{ohm2018versatile} are designed with this requirement in mind. For instance, consider the process of mode selection: the encoder performs multiple calculations to find the mode with the lowest rate-distortion (RD) cost, while the decoder only needs to decode the mode information and perform a single calculation. This asymmetric design allows for efficient content delivery even to devices with constrained processing power.

\begin{figure}[t]
\centerline{\includegraphics[width=3.4in]{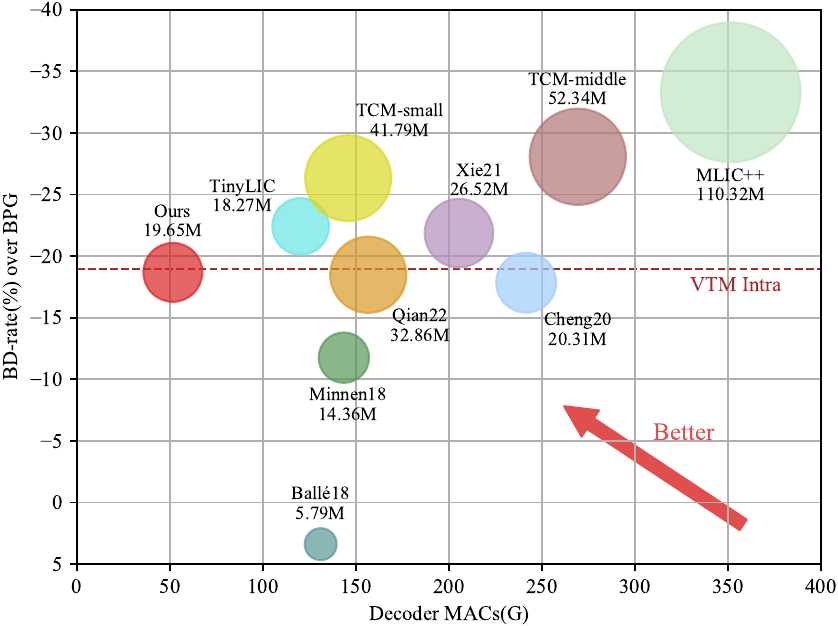}}
\captionsetup{font=small}
\caption{Performance versus decoder complexity on Kodak dataset. Performance is measured by BD-rate against BPG, and decoder complexity metrics include the number of Multiply–Accumulate Operations (MACs) with the input size of $768\times512$ and the number of model parameters. Notable methods like Balle18\cite{balle2018variational}, Minnen18\cite{minnen2018joint}, Cheng20\cite{cheng2020learned}, Xie21\cite{xie2021enhanced}, Qian22\cite{qian2022entroformer}, TinyLIC\cite{lu2022high}, TCM\cite{liu2023learned} and the VVC Intra\cite{ohm2018versatile} are evaluated.
}
\label{comparison}
\vspace{-18pt}
\end{figure}

\begin{figure*}
    \centering
    \captionsetup{font=small}
    \subfloat[Original symmetric structure]{\includegraphics[width=1.81in]{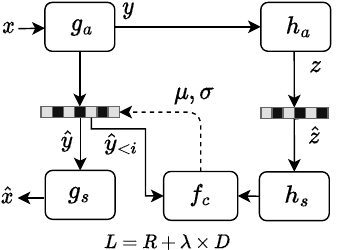}%
    \label{original}}
    \subfloat[Training of synthesis path]{\includegraphics[width=1.81in]{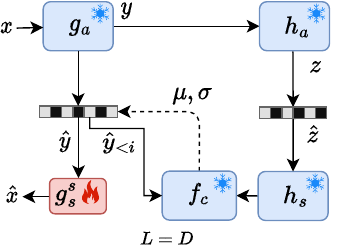}%
    \label{step1}}
    \subfloat[Training of hyperprior path]{\includegraphics[width=1.79in]{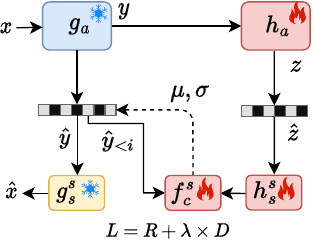}%
    \label{step2}}
    \subfloat[\textbf{Aymmetric lightweight structure}]{\includegraphics[width=1.76in]{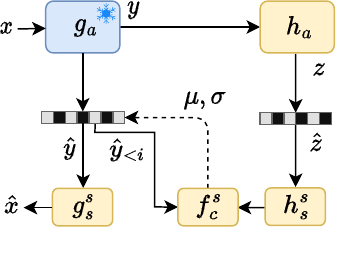}%
    \label{asyml}}
    \captionsetup{font=small}
    \caption{The overall framework of our asymmetric structure and asymmetric training strategy. The snowflake pattern indicates that the module parameters are fixed during training, while the flame pattern indicates that the module parameters are trainable. The yellow boxes indicate that the module parameters have already been altered.}
    \label{train}
    \vspace{-18pt}
\end{figure*}

Learned image compression (LIC) has consistently employed symmetric encoder-decoder architectures since Balle introduced the first LIC method in 2016\cite{balle2016end}. This symmetric design mirrors DCT transforms, where forward and inverse transformations exhibit equivalent computational complexity. Subsequent developments have maintained this design philosophy in pursuit of enhanced performance. Notable advancements include Ball\'e18\cite{balle2018variational}, Minnen18\cite{minnen2018joint,minnen2020channel}, Cheng20\cite{cheng2020learned}, TinyLIC\cite{lu2022high}, and TCM\cite{liu2023learned}. These models, characterized by increasingly complex components, have significantly outperformed traditional algorithms\cite{bao2022nonlinear,bao2023taylor} like VVC intra.

However, the symmetric design in LIC results in comparable encoding and decoding times. As models have grown more complex, the high decoding time poses challenges for practical deployment. While general lightweight design methods exist, such as simplified model structures\cite{howard2017mobilenets} and knowledge distillation\cite{fu2023fast}, the asymmetric demands in compression remain largely unaddressed to our knowledge.

In this paper, we adopt an asymmetric lightweight design for learned image compression architecture, which is called AsymLLIC. We propose a stage-by-stage training strategy that gradually replaces complex decoder modules with simpler ones. Building on this foundation, we explore various efficient decoder architectures. Our experiments demonstrate more efficient decoding while maintaining the same encoding performance. 
This proves that in an LIC system, computational load can be effectively offloaded to the encoding side.

Our contributions can be summarized as follows:

\begin{itemize}
\item An asymmetric encoder-decoder structure and training strategy that maintains high image quality and low bitrate while reducing decoder complexity.

\item Cost-effective designs for synthesis decoder, hyperprior decoder, and context model, optimizing computational complexity and compression performance.

\item Experimental results demonstrating an ideal trade-off between rate-distortion performance and lightweight decoder design. Our model achieves an 18.68\% BD-rate improvement over BPG on Kodak, with only 19.65M parameters and 51.47GMACs in decoder.
\end{itemize}

\section{Related Work}


\subsection{Learned image compression}
Ball\'e et al.\cite{balle2016end} made pioneering work in this field by using neural networks to construct an image compression pipeline comprising an encoder, a decoder, and an entropy model, which was later widely adopted. This approach is based on the transformation coding paradigm, aiming to reduce redundancy through learned transformation. However, the resultant latent space is not independently distributed, leaving spatial redundancy unaddressed. One solution is to get more accurate estimations of the distributions of latent codes. Ball\'e et al.\cite{balle2018variational} proposed a hyperprior to capture the spatial dependencies in the latent code by signaling side information. Following this, more sophisticated entropy models have been introduced to capture these dependencies, such as mean and scale Gaussian distribution\cite{minnen2018joint}, context models\cite{lee2018context}, discretized Gaussian Mixture Likelihoods\cite{cheng2020learned}, and transformer-based entropy models \cite{kim2022joint}\cite{qian2022entroformer}. Another line of research \cite{li2020deep}\cite{cheng2019deep}\cite{xie2021enhanced}\cite{gao2021neural}\cite{mishra2020wavelet}\cite{shin2022expanded}\cite{tang2022joint} focuses on designing more powerful encoders and decoders to better decorrelate the spatial dependencies. For example, Xie et al.\cite{xie2021enhanced} introduced an invertible convolution block for the transformation between images and latent codes. Zou et al.\cite{zou2022devil} proposed a new Transformer-based encoder/decoder that fully leverage both global structure and local texture.

\subsection{Lightweight model design}
While complex network architectures deliver remarkable outcomes, they impose significant computational costs, hindering practical use. Therefore, the pursuit of lightweight model design focuses on optimizing computational efficiency without sacrificing performance. For instance, ShuffleNet \cite{zhang2018shufflenet} enhances performance by introducing channel shuffling to address inter-group information interaction issues in grouped convolutions. Similarly, C-GhostNet \cite{han2020ghostnet} exploits channel redundancy through cost-effective operations, generating similar channel compositions for efficient feature extraction. G-GhostNet \cite{han2022ghostnets} employs a lightweight stage-level network to further enhance feature extraction efficiency. Additionally, some lightweight learned image compression designs have been proposed \cite{he2023efficient}. BG-VAE \cite{zhang2024theoretical} achieves a more efficient model through knowledge distillation, while Qin et al. \cite{qin2024leveraging} accelerate coding speed and reduce complexity by eliminating redundancy between features. However, as overall computational complexity decreases, encoding performance inevitably declines.

\vspace{-3pt}
\section{Proposed Method}
In this section, we first introduce the architecture of learned image compression and propose an asymmetric training strategy to replace decoder modules with simpler ones. We then explore and identify the optimal cost-performance designs for the synthesis decoder, hyperprior decoder, and context model modules, achieving a balance between compression performance and computational overhead.





\begin{figure*}
    \captionsetup{font=small}
    \centering
    \begin{minipage}[b]{0.33\textwidth}
        \centering
        \subcaptionbox{The original structure of synthesis decoder $g_s$\label{fig:a}}{
            \raisebox{0.3cm}
            {\includegraphics[width=\textwidth]{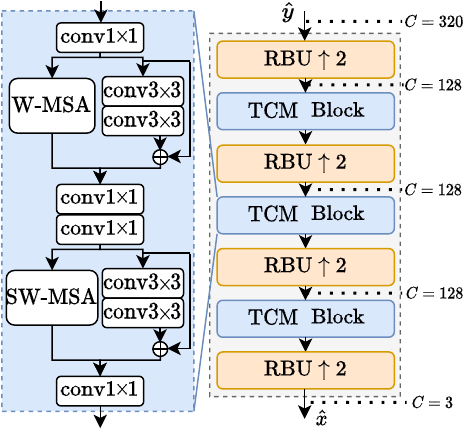}}}
    \end{minipage}%
    \hfill
    \begin{minipage}[b]{0.33\textwidth}
        \centering
        \subcaptionbox{The proposed structure of synthesis decoder $g_s^{s}$\label{fig:b}}{
            \raisebox{0.3cm}
            {\includegraphics[width=\textwidth]{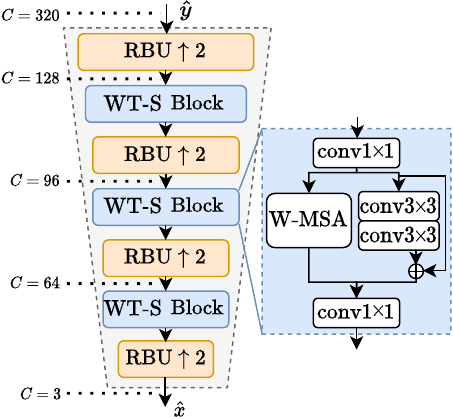}}}
    \end{minipage}%
    \hfill
    \begin{minipage}[b]{0.30\textwidth}
        \centering
        \subcaptionbox{RF of TCM-small\label{fig:c}}{
            \includegraphics[width=0.45\textwidth]{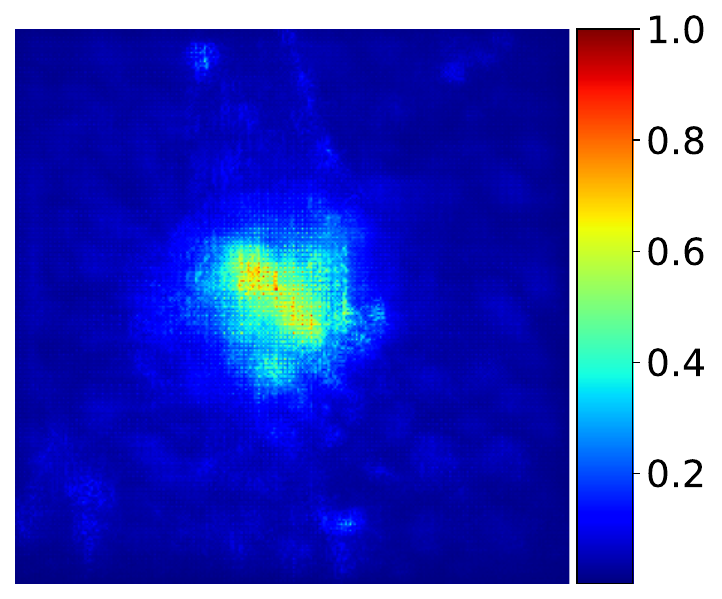}}
        \hfill
        \vspace{0.2cm}
        \subcaptionbox{RF of Ours\label{fig:d}}{
            \includegraphics[width=0.45\textwidth]{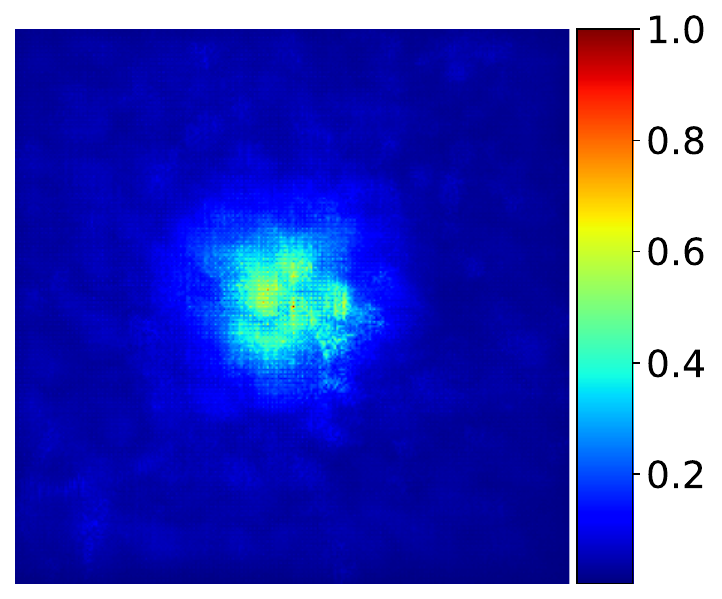}}
        \subcaptionbox{Complexity analysis of synthesis decoder\label{fig:e}}{
            \includegraphics[width=0.95\textwidth]{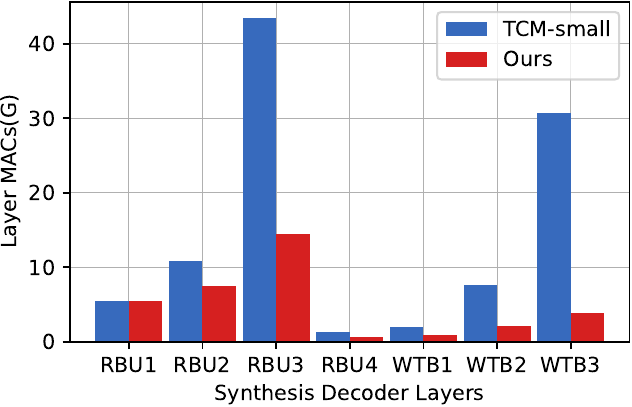}}
    \end{minipage}
    \caption{The designed architecture of synthesis decoder. Subfig (a) and (b) show the original structures of TCM-small and our proposed synthesis decoder. Their receptive fields (RF) are visualized in subfig (c) and (d). Subfig (e) shows the MACs for each layer of the synthesis decoder, where RBU1-4 and WTB1-3 represent the ResidualBlockUpsample layers and transform-based layers from $\hat y$ to $\hat x$.}
    \label{fig:main}
    \vspace{-15pt}
\end{figure*}

\subsection{Asymmetric Training Strategy}\label{3A}

The overall framework of learned image compression is illustrated in Figure \ref{original}. The input image $x$ is mapped into a compact latent representation $y$ through the analysis encoder $g_a$, while the quantized latent representation $\hat{y}$ is inversely mapped to the reconstructed image $\hat{x}$ by the synthesis decoder $g_s$. The hyperprior encoder $h_a$ captures spatial dependency, producing side information $z$. The hyperprior decoder $h_s$ and context model $f_c$ provide the hyperprior parameters for the entropy model. The entire network undergoes end-to-end training with a loss function defined as $L = R + \lambda D$. 
Most learned image compression (LIC) methods employ a symmetric model design, where the structures of $g_a$ and $g_s$, and $h_a$ and $h_s$ are symmetrical. The decoder side, consisting of $g_s$, $h_s$, and $f_c$, is replaced with simpler modules $g^{s}_{s}$, $h^{s}_{s}$, and $f^{s}_{c}$ in our approach.

Designing effective training strategies for this asymmetric architecture is a key challenge. In the first step of our asymmetric training strategy, we fix the parameters of all modules except $g_s$, modify $g_s$ to a lightweight network structure, and train with a loss that includes only the distortion part. This allows for quick fine-tuning to obtain the lightweight $g^{s}_{s}$ model. In the second step, we fix the parameters of all modules except $h_a$, $h_s$, and the context model $f_c$, and perform lightweight modifications on $h^{s}_{s}$ and $f^{s}_{c}$. At this stage, both distortion and rate are included in the training loss. Notably, we do not fix the parameters of $h_a$ because after lightweighting $h_s$, the original hyperprior obtained by $h_a$ cannot adapt to the new $h^{s}_{s}$, thus requiring joint training with the $h_a$ module. Finally, we combine all the required parameters for the decoder to obtain our proposed asymmetric lightweight LIC model, as shown in Figure \ref{asyml}.

\subsection{Lightweight Design of synthesis decoder}\label{3B}

We adopt the network models from TCM-small \cite{liu2023learned}, incorporating the Swin Transformer blocks. The encoder and decoder both use CNN and Transformer blocks stacked with identical channels. We redesign the synthesis decoder $g_s$, the hyperprior decoder $h_s$, and the context network $f_c$ to reduce the decoder-side computational complexity.

\textbf{Swin Transformer Simplification:} In the $g_a$ and $g_s$ modules, the Swin Transformer expands the network's receptive field, leveraging pixel correlations across the entire image to produce a compact latent representation. For the $g_s$ module, we retain only the window-based multi-head self-attention from the Swin Transformer block and remove the subsequent shifted windowing configuration, as shown in Figure \ref{fig:b}. Visualization of the synthesis decoders' receptive fields (RF) in Figures \ref{fig:c} and \ref{fig:d} shows minimal change after removing the shifted windowing configuration. While this approach limits information gathering to a fixed window size, experiments demonstrate that the performance loss is acceptable.

\textbf{Reversed Pyramid Channel Structure:} We design the $g_s$ module with a reversed pyramid channel structure, where the number of channels in the latent representation $y$ gradually narrows while the width and height progressively increase. Figure \ref{fig:e} illustrates the MACs for each layer of the $g_s$ module in both TCM and our method. RBU1-4 and WTB1-3 represent the ResidualBlockUpsample layers and transform-based layers from $\hat{y}$ to $\hat{x}$. This structure prevents the computational complexity of subsequent layers from increasing significantly, maintaining it within a reasonable range. This design ensures effective information utilization while reducing complexity.

\begin{figure*}
    \centering
    \captionsetup{font=small}
    \subfloat[RD performance on Kodak dataset]{\includegraphics[width=2.4in]{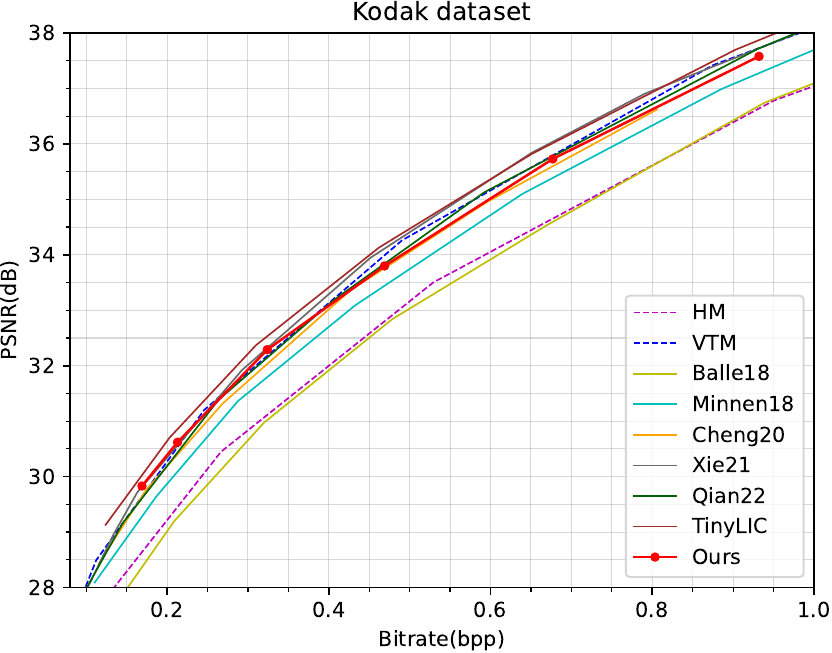}%
    \label{kodak}}
    \subfloat[RD performance on CLIC pro dataset]{\includegraphics[width=2.4in]{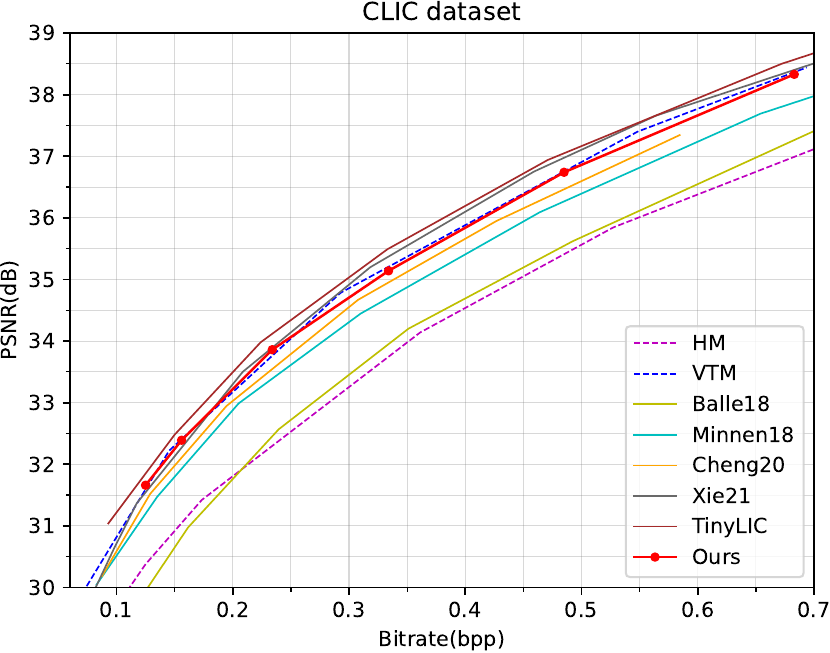}%
    \label{clic}}
    \subfloat[RD performance on Tecnick dataset]{\includegraphics[width=2.4in]{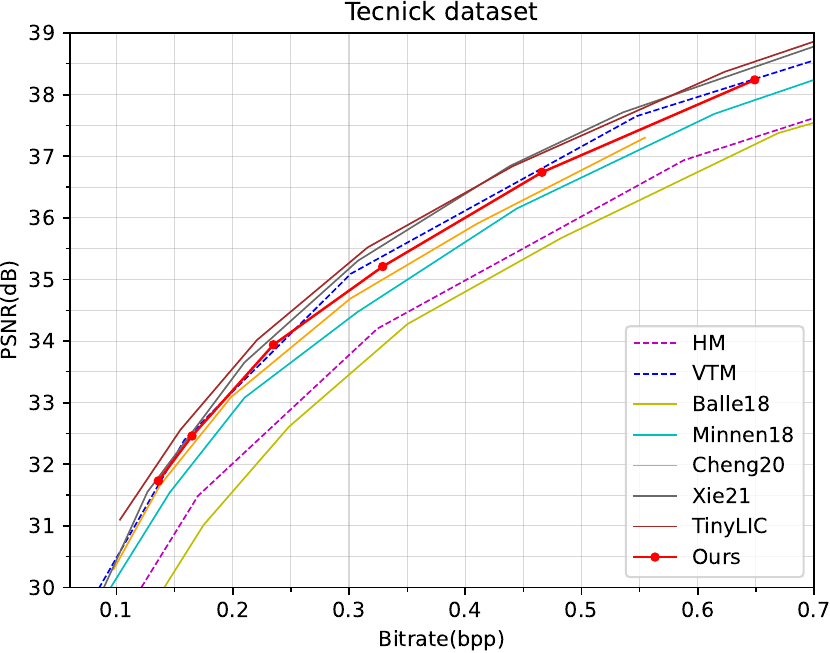}%
    \label{tecnick}}
    \captionsetup{font=small}
    \caption{R-D performance of traditional codec and LIC methods evaluated on different datasets.}
    \label{rdcurve}
    \vspace{-15pt}
\end{figure*}

\subsection{Lightweight Design of hyperprior path}\label{3C}

For the hyperprior path, we adopt the 5-slice channel-wise context model and remove the shifted windowing configuration from the hyperprior decoder and context model. In the context model's slice networks, we further reduce the number of channels in the self-attention module and the number of layers in the residual network. Additionally, we compare the performance impact of changing the number of slices in the hyperprior path while maintaining similar computational complexity. As shown in the experimental results (see Table \ref{hs}), our simplified design results in minimal performance loss.

\section{Experimental Results}
\subsection{Experimental Setting}

\textbf{Training Details:} For training, we randomly choose 300k images
of size larger than 256 × 256 
from ImageNet\cite{deng2009large}, and randomly crop them with the size of 256 × 256 during the training process.
We adopt Adam\cite{kingma2014adam} with a batch size 8 to optimize the network. The initial learning rate is set as $1\times10^{-4}$.
After 0.5M steps, the learning rate is reduced to $1\times10^{-5}$ for the last 0.1M steps.
The model is optimized by RD-loss as $L=R+\lambda D$. Mean square error (MSE) are used to represent the distortion $D$. The $\lambda$ belongs to $\{0.0025, 0.0035, 0.0067, 0.0130, 0.0250, 0.0500\}$.

\textbf{Evaluation:} We test our method on three datasets, i.e., Kodak image set with the image size of $768 \times 512$, old Tecnick test set with the image size of $1200 \times 1200$, CLIC professional validation dataset with up to $2$K resolution. PSNR are used to measure the distortion, while bits per pixel (bpp) are used to evaluate bitrates.

\vspace{-1pt}
\subsection{Rate-Distortion Performance and Complexity Analysis}

From Figure \ref{rdcurve}, it can be observed that, compared to previous LIC models, our model maintains high R-D performance similar to VVC.
Table \ref{rateperformance} presents the BD-rate of our proposed method and other LIC models on the Kodak dataset with BPG (HEVC intra) as anchor. It also shows the total computational complexity (Tot MACs and Tot Parameters) and the decoding complexity (Dec MACs and Dec Parameters). Compared to other methods, our proposed method achieves the lowest decoding and total computational complexity while maintaining high compression performance. Our method strikes an optimal balance between compression performance and computational complexity.


\begin{table}[]
\centering
\caption{
Results of BD-rate (\%) comparsion upon BPG and complexity analysis.
}
\label{rateperformance}
\renewcommand{\arraystretch}{1.2}
\begin{tabular}{>{\centering\arraybackslash}m{0.9cm} >{\centering\arraybackslash}m{0.97cm} >{\centering\arraybackslash}m{1.03cm} >{\centering\arraybackslash}m{1.31cm} >{\centering\arraybackslash}m{0.97cm} >{\centering\arraybackslash}m{1.26cm} }
\hline
        & \textbf{BD-rate} & \textbf{Dec Par.} & \textbf{Dec MACs} & \textbf{Tot Par.} & \textbf{Tot MACs} \\ \hline
Balle18 & 3.38    & 5.79     & 130.96   & 11.82    & 164.34       \\ \hline
Minnen18   & -11.76  & 14.36    & 143.41   & 25.50    & 176.79         \\ \hline
Cheng20 & -17.85  & 20.31    & 241.41   & 29.63     & 403.27        \\ \hline
Xie21   & -21.86  & 26.52    & 205.29   & 50.03     & 407.30         \\ \hline
Qian22  & -18.48  & 32.86    & 156.55   & 44.99     & 193.55         \\ \hline
TinyLIC & -22.41  & 18.27    & 120.18   & 28.34     & 193.41         \\ \hline
TCM     & -26.32  & 41.79    & 145.71   & 44.96     & 211.54         \\ \hline
Ours    & -18.68  & 19.65    & 51.47    & 22.83     & 117.12         \\ \hline
\end{tabular}
\vspace{-15pt}
\end{table}



\subsection{Ablation Study}

\textbf{Discussion on synthesis decoder structure: } 
In the synthesis decoder module, four upsampling processes are necessary to transform the latent representation $\hat y$ into the reconstructed image $\hat x$. We design four different structures of $g_s$ and adjusted parameters to maintain similar MACs. Since the synthesis decoder only affects the PSNR of the reconstructed image, Table \ref{gs} presents the MACs, parameter counts, and average PSNR on the Kodak dataset for different synthesis decoder structures with $\lambda = 0.0067$. Conv\_k5 employs convolutions with a kernel size of 5 for upsampling, RBU utilizes the ResidualBlockUpsample module, TCM\_pruned reduces the number of Swin Transformer blocks to 2 while retaining the shifted windowing configurations, and Ours represents our proposed structure. As shown in Table \ref{gs}, our proposed structure achieves the highest image quality with the second-lowest computational complexity.

\textbf{Discussion on hyperprior path: } 
Similarly, we compare various hyperprior decoder and context model structures. First, we compared the impact of dividing channels into different numbers of slices on compression efficiency in the context model. As shown in Table \ref{hs}, even with similar MACs of hyperprior path, the slice number of 5 achieved significantly lower bit rates compared to structures with fewer slices. TCM\_pruned indicates Swin Transformer blocks with reduced channel numbers while retaining shifted windowing configurations. Table \ref{hs} shows that our proposed structure achieves the lowest bit rate and the second-lowest complexity.

\begin{table}[]
\centering
\caption{
Comparison results of synthesis decoder structures.
}
\label{gs}
\renewcommand{\arraystretch}{1.1}
\setlength{\tabcolsep}{10pt}
\begin{tabular}{cccc}
\hline
\textbf{Structures}  & \textbf{PSNR(dB)}  & \textbf{Params} & \textbf{MACs}  \\ \hline
Conv\_k5  & 31.51 & 1.49   & 30.83 \\ \hline
RBU       & 32.14 & 4.83   & 39.44 \\ \hline
TCM\_pruned       & 31.81 & 3.94   & 45.04 \\ \hline
Ours      & 32.32 & 4.84   & 35.06 \\ \hline
\end{tabular}
\vspace{-5pt}
\end{table}

\begin{table}[]
\centering
\caption{
Comparison results of hyperprior path structures.
}
\label{hs}
\renewcommand{\arraystretch}{1.1}
\setlength{\tabcolsep}{10pt}
\begin{tabular}{cccc}
\hline
\textbf{Structures} & \textbf{Bpp}   & \textbf{Params} & \textbf{MACs}  \\ \hline
slice number = 1 & 0.412 & 20.13  & 9.41  \\ \hline
slice number = 2  & 0.364 & 18.9   & 19.45 \\ \hline
slice number = 5, TCM\_pruned & 0.325 & 16.67  & 19.27 \\ \hline
slice number = 5, Ours  & 0.324 & 14.81  & 16.41 \\ \hline
\end{tabular}
\vspace{-14pt}
\end{table}

\section{Conclusion and Future Work}

In this paper, we investigate the issue of asymmetric design in learned image compression, which is crucial for practical LIC deployment. Specifically, we introduce an asymmetric training strategy that progressively replaces complex decoder modules with simpler ones. Further, we examine the cost-effectiveness of various decoder block structures, evaluating them based on computational complexity and compression performance. Experiments demonstrate that our proposed AsymLLIC requires only 51.47 GMACs of computation and 19.65M decoder parameters to achieve performance comparable to VVC, with significantly lower decoding complexity than previous LIC methods. This implies that in an LIC system, computation load can be effectively offloaded to the encoding side. In the future, we plan to extend the asymmetric design to learned video compression to enable efficient video decoding.

\section{Acknowledgment}
This work was supported by the Fundamental Research Funds for the Central Universities, National Natural Science Foundation of China (62102024, 62331014, 62431015) and Shanghai Key Laboratory of Digital Media Processing and Transmissions, China.

\bibliographystyle{IEEEbib}


\end{document}